\documentclass[aps,twocolumn,floatfix,showpacs,superscriptaddress]{revtex4}
\usepackage{amsmath,amsfonts,amssymb,graphicx,natbib}

\newcommand{\trace}{\mathop{\rm Tr}\nolimits}
\newcommand{\EPR}{\text{EPR}}
\newcommand{\WEPR}{\text{WEPR}}
\newcommand{\GHZ}{\text{GHZ}}

\begin{document}

\title{Multi-particle entanglement under asymptotic
positive partial transpose preserving operations}

\author{S. Ishizaka}
\affiliation{Fundamental and Environmental Research Laboratories,
NEC Corporation, 34 Miyukigaoka, Tsukuba, 305-8501, Japan}
\affiliation{PRESTO, Japan Science and Technology Agency, 4-1-8
Honcho Kawaguchi, 332-0012, Japan}
\author{M.\ B.\ Plenio}%
\affiliation{QOLS, Blackett Laboratory, Imperial College London,
Prince Consort Road, London SW7 2BW, UK}

\affiliation{Institute for Mathematical Sciences, Imperial College
London, 53 Exhibition Road, London SW7 2BW, UK}


\begin{abstract}
We demonstrate that even under positive partial transpose
preserving operations in an asymptotic setting GHZ and $W$ states
are not reversibly interconvertible. We investigate the structure
of minimal reversible entanglement generating set (MREGS) for
tri-partite states under positive partial transpose (ppt)
preserving operations. We demonstrate that the set consisting of
$W$ and EPR states alone cannot be an MREGS. In this context we
prove the surprising result that the relative entropy of
entanglement can be strictly sub-additive for certain pure
tri-partite states which is crucial to keep open the possibility
that the set of GHZ-state and EPR states together constitute an
MREGS under ppt-preserving operations.
\end{abstract}

\date{\today }

\maketitle

{\it Introduction ---} Constraints and resources are intimately
related in physics. If we impose a constraint on a physical
setting then certain tasks become impossible. A resource must be
made available to overcome the restrictions imposed by the
constraints. Such a resource may be manipulated and transformed
under the constrained set of operations but it emerges as a
fundamental law that the resource cannot be created employing only
the constrained set of operations.

An example, motivated by communication scenarios, is to impose the
constraint of local quantum operations and classical communication
(LOCC) in quantum mechanics. The resource that is implied by this
constraint are non-separable states and in particular pure
entangled states such as singlet states, neither of which can be
created by LOCC alone when starting from product states. This
setting gives rise to a theory of entanglement as a resource under
LOCC operations.

Any theory of entanglement as a resource will generally aim to
provide mathematical structures to treat three questions, namely
(1) the characterization of entanglement, (2) the manipulation of
entanglement and (3) the quantification of the entanglement
resource under the given constraint \cite{Plenio V 98,Plenio V
01,Horodecki 01,Horodecki H01,Eisert P 03,Plenio V 05}. Of
particular interest concerning the characterization of
entanglement is the question of how many inequivalent types of
entanglement there are within such a theory. For the theory of
entanglement under LOCC operations we find that in the limit of
infinitely many identically prepared copies of bi-partite pure
states, reversible state transformations can be achieved
\cite{Bennett BPS 96}. As entanglement cannot be created under
LOCC it must remain constant and it is therefore reasonable to say
that the entanglement in different pure bi-partite states is
equivalent, i.e. there is only one kind of pure bi-partite
entanglement. The situation is much less transparent in the
multi-particle setting. It has been proven in the tri-partite
setting that for example Greenberger-Horne-Zeiliner (GHZ) states
\begin{equation}
    |\GHZ\rangle = \frac{|000\rangle + |111\rangle}{\sqrt{2}}
\end{equation}
and Einstein-Podolsky-Rosen (EPR) states
\begin{equation}
    |\EPR\rangle = \frac{|01\rangle - |10\rangle}{\sqrt{2}}
\end{equation}
cannot be reversibly interconverted even in the asymptotic limit
\cite{Linden PSW 99}. The question whether all other tri-partite
pure states can be created reversibly from GHZ and EPR states has
also been considered and answered negatively \cite{Galvao PV
00,Acin VC 02}. The existence and structure of the smallest set of
pure tri-partite states that allows for the reversible generation
of all other tri-partite pure states, a so-called minimal
reversible entanglement generating set (MREGS) \cite{Bennett PRST
99}, is still undecided and it is possible that such a set must
contain infinitely many elements.

A different setting for an entanglement theory is motivated by the
concept of partial time reversal or partial transposition. For two
qubits, density matrices that remain positive under partial
transposition (denoted as ppt-states) are exactly the separable
states. For higher dimensions this is not the case as there are
ppt-states that are inseparable \cite{Horodecki HH 98}. The
positivity under partial transposition motivates the definition of
the set of positive-partial-transpose-preserving operations
(ppt-operations) which are defined as those operations that map
all ppt-states to ppt-states \cite{Rains 01}. In this setting, the
implied resource are all those states that are not positive under
partial transposition.

The theory of entanglement under ppt-operations still possesses
the property that in an asymptotic setting pure state
transformations are reversible and that consequently there is only
one type of pure state entanglement. The additional power afforded
by ppt-operations as compared to LOCC operations becomes apparent
both in the mixed state and the multi-party setting. In the mixed
state setting examples for reversible state transformations have
been discovered \cite{Audenaert PE 03} and the possibility remains
open that in this setting all entanglement reduces to only one
type, in stark contrast to the LOCC setting where reversible mixed
entanglement transformations are known only in trivial cases
\cite{Vidal C 02, Horodecki SS 02}. In the non-asymptotic setting
for pure state it has been shown that both under ppt-operations
\cite{Audenaert PE 03} and under LOCC operations supported by
ppt-bound entanglement \cite{Ishizaka 04} state transformations
become possible that are impossible under LOCC. Indeed, it has
been shown in \cite{Ishizaka P 04} that on the single copy level
we can use trace preserving completely positive ppt-operations to
transform for example a GHZ state
into a $W$-state
\begin{equation}
    |W\rangle = \frac{|001\rangle + |010\rangle + |100\rangle}{\sqrt{3}}
\end{equation}
with a maximal success probability of
\begin{displaymath}
    p = \frac{1}{4}(-2+(18-6\sqrt{3})^{1/3}+(18+6\sqrt{3})^{1/3})\approx
    0.75...
\end{displaymath}
This is remarkable in particular because under LOCC this
transformation has a strictly zero success probability \cite{Dur
VC 00}. The surprisingly large success probability and the proven
existence of some asymptotically reversible state transformations
under ppt-operations in the bi-partite setting suggests that a
theory of entanglement under ppt-operations might have a simpler
structure than that under the LOCC constraint. Motivated by this
we will consider the MREGS problem under the more general setting
of ppt-preserving operations. We will show that GHZ and $W$ state
remain asymptotically inequivalent even under ppt-operations. We
then explore possible ppt-MREGS and find that the set consisting
of the GHZ and the EPR state is a very promising example for an
MREGS under ppt-preserving operations.

{\it Asymptotic irreversibility between GHZ and $W$ state ---} To
demonstrate the asymptotic irreversibility between GHZ and $W$ under
ppt-operations we will prove that there is one entanglement
measure under which GHZ is strictly more entangled than the $W$
state and another entanglement measure where this relation is
reversed. Such an example then provides a contradiction to the
invariance of entanglement measures under reversible state
transformations. As we are considering an asymptotic setting,
there are two constraints to be observed in the choice of
entanglement measures. Firstly, we have to employ the appropriate
asymptotic versions $E^{\infty}$ of entanglement measures $E$
which are defined as
\begin{equation}
    E^{\infty}(\sigma_{ABC}) = \lim_{n\rightarrow\infty}
    \frac{E(\sigma_{ABC}^{\otimes})}{n}
\end{equation}
Furthermore, we require that these asymptotic measures are
continuous, as we must include in our considerations situations
where the final state is reached asymptotically as a limit of
progressively more similar states \cite{Audenaert PE 03}. These
requirements limit the choice of available entanglement measures
quite severely.

The first two measures are based on the relative entropy of
entanglement \cite{Relent,Vedral P98} with respect to an appropriately
defined set of states ${\cal T}$, which is given by
\begin{equation}
    E(\sigma) = \inf_{\rho\in{\cal T}} S(\sigma||\rho)
\end{equation}
where $S(\sigma||\rho)=\trace\{\sigma\log_2\sigma-\sigma\log_2\rho\}$.

First, we consider the asymptotic two-party relative entropy of
entanglement $E_{A:BC}(\sigma)$ across all the possible bi-partite
splits of the three parties. For pure states this measure can
actually be computed analytically and equals the entropy of
entanglement \cite{Bennett BPS 96}, i.e. the entropy of the reduced
density operator of one party. This measure is asymptotically
continuous \cite{Donald H 99}. We then find
\begin{equation}
    E_{A:BC}(\sigma_W) = \log_2 3 - \frac{2}{3} < 1 = E_{A:BC}(\sigma_{\GHZ})
    \label{eq: entropy}
\end{equation}
where $\sigma_W = |W\rangle\langle W|$ and
$\sigma_{\GHZ}=|\GHZ\rangle\langle \GHZ|$.

Secondly, we consider the tri-partite relative entropy of
entanglement with respect to states that are ppt for every
bi-partite split \cite{Plenio V 01a}.
\begin{equation}
    E_{ABC}(\sigma_{ABC})=\inf_{\rho_{ABC} \in \hbox{\scriptsize tri-ppt}}
    S(\sigma_{ABC}||\rho_{ABC}),
\end{equation}
Generally, the asymptotic relative entropy of entanglement is very
difficult to compute except in situations with high symmetries
\cite{Audenaert EJPVD 01}, but useful lower bounds exist. Indeed,
Theorem 1 in \cite{Plenio V 01a} states that
\begin{equation}
    E^\infty_{ABC}(\sigma_{ABC})\ge
    \max_{i=(AB),(AC),(BC)}
    \{E_i^\infty(\sigma_{i})+ S(\sigma_{i})\}
    \label{eq: lower bound}
\end{equation}
where $\sigma_{AB}=\trace_C\sigma_{ABC}$ and $E_i$ is the relative
entropy of entanglement for bi-partite states. We then find
\begin{equation}
    E^\infty_{ABC}(\sigma_{\GHZ}) = 1.
    \label{ghz}
\end{equation}
The final step of the proof consists in the verification of
$E_{ABC}^{\infty}(\sigma_W)>1$. It is known \cite{Plenio V 01a,Wei EGM 04} that
    $E_{ABC}(\sigma_W)=\log_2 9/4$
but this is not sufficient as $E_{ABC}$ can be strictly
sub-additive for arbitrary pure tri-partite states. Indeed,

{Lemma:} {\it The relative entropy of entanglement $E_{ABC}$ for
the pure state (\ref{antisymmetric}) is strictly sub-additive.}

{\em Proof:}
For the spin-$0$ state for a tri-partite state of three level systems
\begin{equation}
    |A\rangle = \frac{1}{\sqrt{6}}\sum_{i=1}^{3}
    \epsilon_{ijk}|i\rangle|j\rangle|k\rangle\label{antisymmetric}
\end{equation}
where $\epsilon_{ijk}$ is the totally anti-symmetric tensor,
we find $E_{ABC}(|A\rangle\langle
A|)=\log 6$, both with respect to LOCC states and with respect to
ppt-states. To demonstrate strict sub-additivity of the
tri-partite relative entropy of entanglement we now consider the
relative entropy of entanglement for two copies of
$|A\rangle^{\otimes 2}$ given in Eq. (\ref{antisymmetric}). To
this end we will provide a guess for a closest ppt-state. We
define the product state $|m\rangle =
|m_A\rangle|m_B\rangle|m_C\rangle$ with
$|m_A\rangle=|m_B\rangle=|m_C\rangle=(|00\rangle+|11\rangle+|22\rangle)/\sqrt{3}$.
Now we employ the invariance of $|A\rangle\langle A|$ under the
application of $U\otimes U\otimes U$, i.e. twirling. Twirling the
state $|m\rangle\langle m|$ leaves its overlap with
$(|A\rangle\langle A|)^{\otimes 2}$ invariant so that we then find
that the resulting state $\rho$ is of the form $\rho = F
(|A\rangle\langle A|)^{\otimes 2} + \rho_{\perp}$ where $F=1/27$
and $\rho_{\perp}$ is orthogonal on $(|A\rangle\langle
A|)^{\otimes 2}$. This state is certainly separable and therefore
ppt. As a consequence we find
\begin{eqnarray*}
    E_{ABC}((|A\rangle\langle A|)^{\otimes 2})
    &\le & S((|A\rangle\langle A|)^{\otimes 2}||\rho)\\
    &=& -tr(|A\rangle\langle A|)^{\otimes 2}\log \rho\\
    &=& \log\, 27\\
    &<& 2\log 6\\
    &=& 2 E_{ABC}(|A\rangle\langle A|))\, .
\end{eqnarray*}
This completes the proof of the Lemma.\\

Straightforward additive lower bounds on the bi-partite relative
entropy of entanglement \cite{Plenio VP 00} such as
\begin{displaymath}
    E_{AB}(\sigma_{AB})\ge \max\{S(\sigma_A),S(\sigma_B)\}-
    S(\sigma_{AB})
\end{displaymath}
are not sufficient either to prove $E_{ABC}^{\infty}(\rho_W)>1$.
We can, however, employ the distillable entanglement to lower
bound the bi-partite relative entropy of entanglement for the
state $\trace_A \rho_W$. In section III.B.2 of \cite{Bennett DSW 96} a
specific distillation employing two-way communication has been
presented which, for the state $\trace_A \rho_W$, yields an asymptotic
gain of 
$D(\trace_A \rho_W) = 1/9$
of an EPR pair per copy. While the efficiency of this procedure
can be increased \cite{Distillation} it is sufficient for our
present purpose as it provides a lower bound on the asymptotic
rrelative entropy of entanglement. With Eq. (\ref{eq: lower
bound}) and Eq. (\ref{ghz}) we then find
\begin{eqnarray}
    E^\infty_{ABC}(\sigma_W) &\ge&
    \frac{1}{9} - \frac{1}{3}\log_2\frac{1}{3}
    - \frac{2}{3}\log_2\frac{2}{3} \nonumber\\ [0.2cm]
    &>& 1 = E^\infty_{ABC}(\sigma_{\GHZ}) .
    \label{eq: W}
\end{eqnarray}
As a consequence, Eqs. (\ref{eq: entropy}) and (\ref{eq: W}) and
the continuity of the measures that we have employed demonstrate
that it is impossible to satisfy both
\begin{eqnarray*}
    E^\infty_{ABC}(\sigma_{\GHZ}^{\otimes n_{\GHZ}})&=&
    E^\infty_{ABC}(\sigma_W^{\otimes n_W}),\\ [0.2cm]
    E^\infty_{A:BC}(\sigma_{\GHZ}^{\otimes n_{\GHZ}})&=&
    E^\infty_{A:BC}(\sigma_W^{\otimes n_W})
\end{eqnarray*}
for any choice of $n_W$ and $n_{\GHZ}$. Consequently, the
asymptotic reversible transformation between GHZ and $W$ is
impossible even under ppt-operations.
\par
{\it Exploration of ppt-MREGS ---} The result proven above implies
that the neither the set $\{\GHZ\}$ nor $\{W\}$ form a ppt-MREGS.
In the following we will explore whether the addition of a set of
EPR states $\{\EPR_{AB}, \EPR_{AC}, \EPR_{BC}\}$ can suffice to create an ppt-MREGS.

Let us begin by considering the set $\{W, \EPR_{AB}, \EPR_{AC}, \EPR_{BC}\}$
and explore whether it is possible to obtain a GHZ state reversibly from
this set. Let us denote
\begin{displaymath}
    \sigma_{\WEPR}=\sigma_W^{\otimes n_W}\otimes \EPR_{AB}^{\otimes n_{AB}} \otimes
    \EPR_{AC}^{\otimes n_{AC}}\otimes \EPR_{BC}^{\otimes n_{BC}}
\end{displaymath}
with positive $n_W, n_{AB}, n_{AC}, n_{BC}$. Reversibility then
requires that
\begin{eqnarray}
    E_{ABC} (\sigma_{\WEPR}) &=& E_{ABC}(\sigma_{\GHZ}^{\otimes n_{\GHZ}}),
    \label{eq: ABC}\\ [0.15cm]
    E_{A:BC}(\sigma_{\WEPR}) &=& E_{A:BC}(\sigma_{\GHZ}^{\otimes n_{\GHZ}}),
    \label{eq: A:BC}\\ [0.15cm]
    E_{B:AC}(\sigma_{\WEPR}) &=& E_{B:AC}(\sigma_{\GHZ}^{\otimes n_{\GHZ}}),
    \label{eq: B:AC}\\ [0.15cm]
    E_{C:AB}(\sigma_{\WEPR}) &=& E_{C:AB}(\sigma_{\GHZ}^{\otimes n_{\GHZ}})
    \label{eq: C:AB}
\end{eqnarray}
are satisfied. As a consequence of the permutation symmetry of the
$W$ and GHZ-state we find that
$n_{AB}\!=\!n_{AC}\!=\!n_{BC}\!=\!n$ and in the asymptotic limit
Eqs. (\ref{eq: A:BC}) - (\ref{eq: C:AB}) all lead to
\begin{equation}
    n_W H(\frac{1}{3})+2n=n_{\GHZ}.
    \label{eq: WEPR1}
\end{equation}
Due to the lack of general additivity of $E_{ABC}$ we need to
employ in Eq.\ (\ref{eq: ABC}) the bound of Eq.\ (\ref{eq: lower
bound}) as well as the additivity of
$E^\infty_2(\rho\otimes\sigma)\!=\!E^\infty_2(\rho)\!+\!E^\infty_2(\sigma)$
for a pure as well as separable $\sigma$ \cite{Rains 01}. Then we
find
\begin{equation}
    n_W (E^\infty_2(\trace_A W)+H(\frac{1}{3}))+3n \le n_{\GHZ}.
    \label{eq: WEPR2}
\end{equation}
Employing again that $E^\infty_2(\trace_A W)\!\ge\!\frac{1}{9}$
we find that Eqs. (\ref{eq: WEPR1}) and (\ref{eq: WEPR2}) can only
be satisfied when $n\!<\!0$ which is a contradiction.

The above result suggests that a more promising choice for the
ppt-MREGS is $\{\GHZ, \EPR_{AB}, \EPR_{AC}, \EPR_{BC}\}$. Let us now
consider whether it is possible to obtain the $W$-state reversibly
from this set. Let us denote
\begin{displaymath}
    \sigma_G=\sigma_{\GHZ}^{\otimes n_{\GHZ}}\otimes \EPR_{AB}^{\otimes n_{AB}} \otimes
    \EPR_{AC}^{\otimes n_{AC}}\otimes \EPR_{BC}^{\otimes n_{BC}}.
\end{displaymath}
with positive $n_{AB}, n_{AC}, n_{BC}$ and $n_{\GHZ}$. Reversibility
then implies
\begin{eqnarray}
    E_{ABC}(\sigma_G) &=& E_{ABC}(\sigma_W^{\otimes n_{W}}),
    \label{eqI: ABC} \\ [0.15cm]
    E_{A:BC}(\sigma_G)&=& E_{A:BC}(\sigma_W^{\otimes n_{W}}),
    \label{eqI: A:BC}\\ [0.15cm]
    E_{B:AC}(\sigma_G) &=& E_{B:AC}(\sigma_W^{\otimes n_{W}}),
    \label{eqI: B:AC}\\ [0.15cm]
    E_{C:AB}(\sigma_G) &=& E_{C:AB}(\sigma_W^{\otimes n_{W}}).
    \label{eqI: C:AB}
\end{eqnarray}
Again, Eqs.\ (\ref{eqI: A:BC}) - (\ref{eqI: C:AB}), imply
$n_{AB}\!=\!n_{AC}\!=\!n_{BC}\!=\!n$. Due to the additivity of
$E_{ABC}$ for the state $\sigma_G$ we find in the asymptotic limit the
equations
\begin{eqnarray*}
    n_{\GHZ} + 3n &=& n_W E^{\infty}_{ABC}(\sigma_W),\\
    n_{\GHZ} + 2n &=& n_W H(\frac{1}{3}).
\end{eqnarray*}
These two equations are solved by
\begin{eqnarray*}
    \frac{n}{n_W} &=& E^{\infty}_{ABC}(\sigma_W)-
    H(\frac{1}{3}) > 0\\
    \frac{n_{\GHZ}}{n_W} &=& 3H(\frac{1}{3}) -
    2E^{\infty}_{ABC}(\sigma_W) > 0.
\end{eqnarray*}

We are therefore unable to rule out that a $W$-state can be
generated asymptotically reversibly from $\{\GHZ,
\EPR_{AB}, \EPR_{AC}, \EPR_{BC}\}$.
In general, using a set of constraints analogous to Eqs.\
(\ref{eqI: ABC})- (\ref{eqI: C:AB}), it is found that
necessary constraints so that a tripartite state $|\psi\rangle$
can be generated reversibly from the set are
\begin{equation}
\max\{E_{A:BC}(\psi),E_{B:AC}(\psi),E_{C:AB}(\psi)\}\!\le\!E^\infty_{ABC}(\psi)
\label{eq: lower condition}
\end{equation}
and
\begin{equation}
E^\infty_{ABC}(\psi)
\!\le\! \frac{E_{A:BC}(\psi)+E_{B:AC}(\psi)+E_{C:AB}(\psi)}{2}.
\label{eq: upper condition}
\end{equation}
The former constraint Eq.\ (\ref{eq: lower condition}) is satisfied for
an arbitrary $|\psi\rangle$ because of the lower bound for
$E^\infty_{ABC}(\psi)$ of Eq.\ (\ref{eq: lower bound}).
For the $W$ state, the latter constraint of
Eq.\ (\ref{eq: upper condition}) can be satisfied even if
the relative entropy of entanglement is additive
because
\begin{equation}
E^\infty_{ABC}(\sigma_W)\le E_{ABC}(\sigma_W) < \frac{3}{2}H(\frac{1}{3}).
\end{equation}
For some tripartite pure states however the subadditivity of the relative
entropy of entanglement plays a crucial role for the satisfaction
of Eq.\ (\ref{eq: upper condition}) as shown below.

%
The spin-$0$ state given by Eq. (\ref{antisymmetric})
exhibits strict sub-additivity of the relative entropy of
entanglement as had been shown in the Lemma.
From \cite{Plenio V 01a} we find a lower bound of $\log_2 5$ on
the asymptotic relative entropy of entanglement with respect to
ppt-states.
The constraint Eq.\ (\ref{eq: upper condition}) requires that $E^{\infty}_{ABC}(A)\le
\frac{1}{2}\log 27 \approx 1.02\log 5$. While we do not know the
asymptotic relative entropy of entanglement $E^{\infty}_{ABC}(A)$
precisely we can bound it from above by providing an upper bound
on $E(A^{\otimes 2})$ which in turn yields an upper bound on
$E^{\infty}_{ABC}(A)$. We find that $E(A^{\otimes 2})\le \log 27$
which again shows that we cannot rule out that $\{\GHZ,
\EPR_{AB}, \EPR_{AC}, \EPR_{BC}\}$ is an ppt-MREGS.

{\it Conclusion ---} In the single copy setting for tri-partite
entanglement it can be shown that a single copy of a GHZ state can
be converted to a $W$ state with a success probability of more than
75\% employing trace-preserving ppt-operations. Motivated by this
we have investigated the manipulation of tri-partite states with
ppt-preserving operations in the asymptotic setting. The large
success probability for the GHZ to $W$ transformation in the single
copy setting could have pointed to a possible asymptotic
reversibility of these states or, at least, a simpler structure of
MREGS under ppt-preserving operations. Despite this large success
probability, however, we have been able to demonstrate that even
in an asymptotic setting GHZ and $W$ state do not become reversibly
interconvertible. We furthermore explored the minimal size of a
MREGS under ppt-preserving operations. We have been able to prove
that the set consisting of $W$ and EPR states alone cannot form
an MREGS. The set of GHZ state and EPR states however
constitutes a promising candidate for an MREGS under
ppt-preserving operations. This result suggest that by comparison
to the situation under LOCC operations the structure of MREGS
under ppt-preserving transformations might be simplified.

\par
%
{\it Acknowledgments ---} We are grateful to K. Audenaert for his
help in finding the example provided in the proof of the Lemma. We
also acknowledge discussions with B. Groisman, N. Linden, S.
Popescu and A. Winter on the issues discussed here. This research
was initiated during two visits to the ERATO project on {\it
Quantum Computation and Information}. This work is part of the
QIP-IRC (www.qipirc.org) supported by EPSRC (GR/S82176/0) as well
as the EU (IST-2001-38877), a Royal Society Leverhulme Trust
Senior
Research Fellowship and the Leverhulme Trust.\\


\begin{thebibliography}{99}
\bibitem{Distillation} We have been able to increase the yield to
approximately $0.17$. The procedure is unlikely to be optimal and
it is conceivable that one can achieve $E_D(tr_A\rho_W)=log_2 3 -
\frac{4}{3}$ which would guarantee the additivity of $E_{ABC}$ for
the $W$-state.
\end{thebibliography}
\end{document}